# *Water evaporation-driven dynamic diode for direct electricity generation*

*Jiarui Guo[1], Xuanzhang Hao[2], Yuxia Yang[1], Shaoqi Huang[1], Zhihao Qian[1], Minhui Yang[1], Hanming Wu[3], Liangti Qu[2*], Novoselov Kostya S[4*], Shisheng Lin[1,5*]*

[1]College of Information Science and Electronic Engineering, Zhejiang University, Hangzhou, 310027, P. R. China.

[2]Laboratory of Flexible Electronics Technology, Key Laboratory of Organic Optoelectronics & Molecular Engineering, Ministry of Education, Department of Chemistry, Tsinghua University, Beijing 100084, P. R. China.

[3]College of Integrated Circuits, Zhejiang University, Hangzhou, 311200, P. R. China.

[4]Natl Univ Singapore, Dept Mat Sci & Engn, 03-09 EA, Singapore, Singapore.

[5]State Key Laboratory of Modern Optical Instrumentation, Zhejiang University, Hangzhou, 310027, P. R. China.

*Corresponding author. Email: shishenglin@zju.edu.cn ;lqu@mail.tsinghua.edu.cn; msekst@nus.edu.sg.

## Abstract

Harnessing energy from ubiquitous water resources via molecular-scale mechanisms remains a critical frontier in sustainable energy research. Herein, we present a novel evaporation-driven power generator based on a dynamic diode architecture that continuously harvests direct current (DC) electricity by leveraging the flipping of the strong built-in electric field (up to $10^{10}$ V/cm) generated by polar molecules such as water to drive directional carrier migration. In our system, water molecules undergo sequential polarization and depolarization at the graphene–water–silicon interface, triggering cycles of charge trapping and release. This nonionic mechanism is driven primarily by the Fermi level difference between graphene and silicon, augmented by the intrinsic dipole moment of water molecules. Structural optimization using graphene enhances evaporation kinetics

**and interfacial contact, yielding an open-circuit voltage of 0.35 V from a 2 cm × 1 cm device. When four units are connected in series, the system delivers a stable 1.2V output. Unlike ion-mediated energy harvesters, this corrosion-free architecture ensures long-term stability and material compatibility. Our work introduces a fundamentally new approach to water-based power generation, establishing interfacial polarization engineering as a scalable strategy for low-cost, sustainable electricity production from ambient water.**

## Introduction

Water, as Earth's most abundant and dynamic substance, plays a central role in driving global energy cycles through continuous phase transitions such as evaporation and condensation.[1] These processes, fueled by solar radiation, mediate massive transfers of heat and mass across the biosphere and atmosphere.[2] Despite the ubiquity and renewability of this natural hydrological cycle, harnessing its intrinsic energy—particularly from the molecular behavior of water during evaporation—remains an underexplored opportunity in sustainable energy science.[3,4]

Direct current electricity serves as the lifeblood of critical modern technologies—powering fault-tolerant telecommunication grids with 48V/60A backup systems, enabling efficient green hydrogen production through direct wind-to-electrolyzer coupling, and sustaining aerospace systems in extreme environments.[5–7] However, conventional DC generators suffer from structural complexity, high production costs, demanding maintenance requirements, and reliability issues.[8–10] Traditional dynamic diodes face additional constraints, limited application scenarios and short operational lifespans, hindering their widespread adoption.[11–14] Hydrovoltaic energy conversion has gained significant attention as an emerging technology, offering unique benefits such as simple device architecture and access to abundant natural energy sources.[15] Water evaporation, a ubiquitous process in nature, serves as a key mechanism for energy harvesting.[16–18] Early work by Guo group demonstrated electricity generation from water evaporation in carbon black, revealing the potential of evaporation-induced electrical potentials.[19,20] Conventional evaporation-based systems, in contrast, required complex fabrication and high-performance materials.[21–24]

We break this paradigm through molecular interfacial engineering and pioneer the first demonstration of semiconductor DC generation driven by molecular-scale interfacial dynamics—termed the "dynamic diode" mechanism. The graphene-water-semiconductor dynamic diode exploits evaporation-triggered alignment of water dipoles and subsequent depolarization cycles, leveraging the flipping of the strong built-in electric field (up to $10^{10}$ V/cm) from polar molecules such as water to drive directional carrier migration, thereby driving sustained unidirectional electron flow. This mechanism operates independently of fluid motion direction—a radical departure from existing platforms—while circumventing electrode corrosion endemic to ionic systems. Our advances establish three pivotal milestones: universal voltage programmability via Fermi level control, scalable outputs reaching electronics-compatible 1.4 V through modular integration, and synergistic solar-energy harvesting amplifying performance to 0.75 V/1.05 µA under 50 W/m² illumination. The operational scope extends beyond water to ethanol/acetone, confirming non-electrolyte molecular polarization as the core engine.

## Results and Discussion

**Device Architecture and Interfacial Polarization Dynamics.** Figure 1a shows the device structure comprising an n-type silicon substrate (<100> orientation, 0.01 Ω·cm resistivity) with a Ti/Au back electrode as the bottom charge-collecting layer. Monolayer graphene transferred onto polyethylene terephthalate (PET) forms the top charge-collecting layer. As illustrated in the cross-section (Figure 1b), water molecules polarize at the interfaces due to the Fermi level difference, confining electrons at silicon and holes at graphene surfaces. During evaporation, movement of the three-phase contact line triggers molecular depolarization. This interfacial process enables continuous charge accumulation and release, generating sustained current (Figure 1c). The system delivers 80 nA current and ≈0.35 V voltage per cm² (Figures 1d-e).

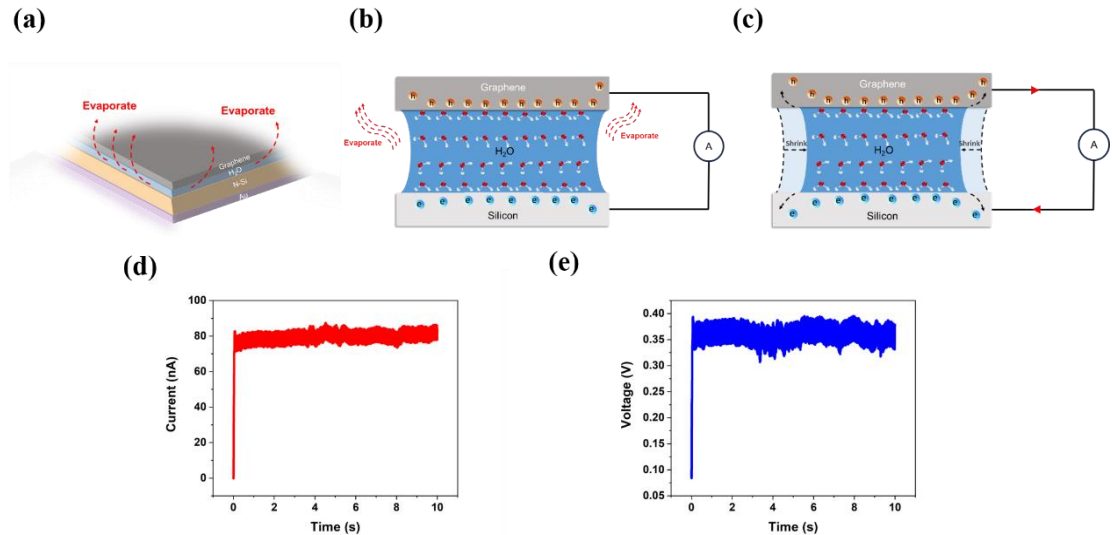

**Fig. 1 | Schematic illustrations of the graphene-water-silicon dynamic diode device driven by water evaporation.** (a) Schematic diagram of the device architecture. (b) Evaporation dynamics of water within the device. (c) Electricity generation mechanism during liquid film contraction induced by evaporation. (d) Current output characteristics of the device. (e) Voltage generation profile of the device.

**Mechanistic Insights: Polarization-Driven Charge Generation.** The DC output follows a polarization-depolarization cycle governed by three key factors: Fermi level difference ($\Delta E_F$), liquid evaporation kinetics, and interfacial dipole orientation. Initially, water molecules align uniformly under $\Delta E_F$, establishing electrostatic equilibrium and forming opposing charge reservoirs ($\pm Q_0$) at the interfaces (Figure 2a). During evaporation, liquid film contraction disrupts dipole alignment, triggering depolarization and carrier release (Figure 2b). Systematic experiments with metal-water-silicon heterostructures confirm that both sign and magnitude of $\Delta E_F$ dictate output characteristics.[25,26] Under controlled conditions (25°C, 45% RH), silicon (work function: 4.34 eV) paired with various electrodes—graphene (4.60 eV), Cu (4.48 eV), Ag (4.26 eV), and Al (4.28 eV)—yields voltages scaling linearly with $\Delta E_F$ (Figures 2c-d). Positive $\Delta E_F$ configurations (graphene/Si: +0.26 eV; Cu/Si: +0.14 eV) generate forward voltages of 0.4 V and 0.2 V, while negative $\Delta E_F$ pairs (Ag/Si: -0.08 eV; Al/Si: -0.06 eV) produce reverse voltages of -0.18 V and -0.3 V (30 μL droplet). Notably, the native $Al_2O_3$ layer on aluminum enhances effective $\Delta E_F$, boosting output by ~18% versus theoretical predictions (Figure 2d).

The essential role of molecular dipole alignment is validated via comparative testing with polar and nonpolar liquids (Figure 2e). Polar non-electrolytes such as ethanol (0.30 V), acetone (0.26 V), and isopropanol (0.24 V) show reduced outputs relative to water, correlating with their dipole moments (1.69 D → 0.18 D). Nonpolar liquids (e.g., $CCl_4$, $C_6H_{14}$) yield no measurable response. This voltage hierarchy ($H_2O$ > EtOH > Acetone > IPA) is governed by both dipole magnitude and hydrogen bonding structure. Water's tetrahedral network (coordination number ≈3.8) maintains persistent polarization, outperforming weaker, less ordered alcohol systems. The introduction of electrolytes like NaCl into water attenuates the device's output voltage from ~0.3 V at 0.1 mol/L to ~0.05 V at 1.0 mol/L (Figure 2f). This reduction arises due to disruption of water dipole alignment by $Na^+/Cl^-$ ions and competitive adsorption at interfaces, fundamentally impairing polarization-driven charge generation.

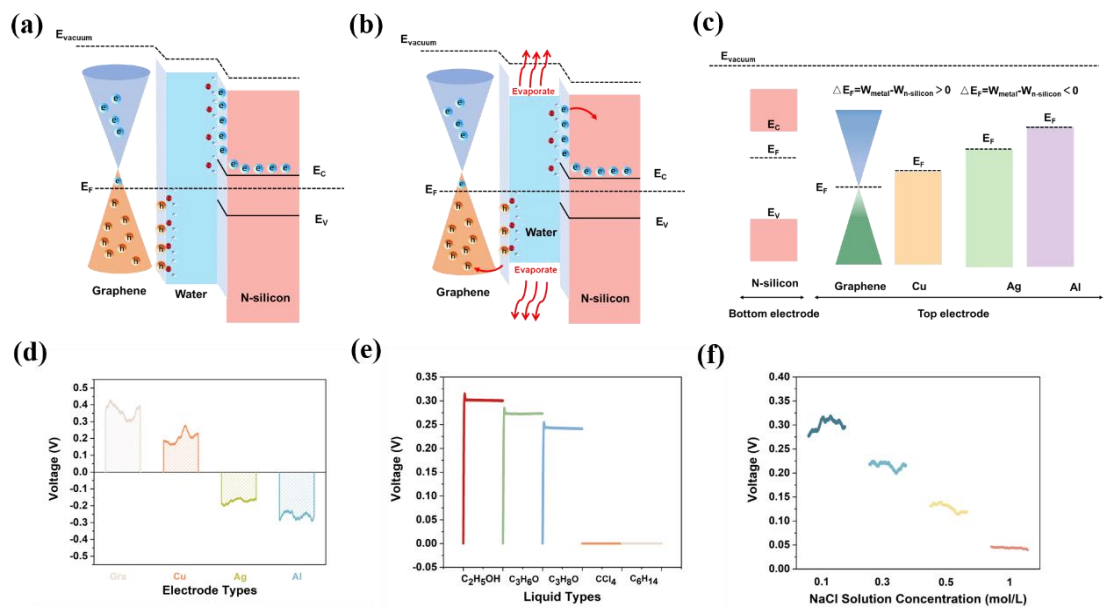

**Fig. 2 | Theoretical framework based on Fermi level difference(ΔEF) and molecular polarization.** (a) Energy band diagram prior to water molecule evaporation. (b) Charge carrier dynamics during water evaporation. (c) Simplified electronic band alignment diagrams for various metals (Ag, Al, Cu) and n-type silicon. (d) Output voltage profiles of water-evaporation-driven devices with metallic electrodes (2 cm × 1 cm active area). (e) Voltage generation across polar liquids ($C_2H_5OH$, $C_3H_6O$, $C_3H_8O$) and nonpolar liquids ($CCl_4$, $C_6H_{14}$). (f) Output voltage dependence on NaCl concentration (0-1.0 M) in dynamic graphene-electrolyte-silicon system. All experiments conducted with 30 μL droplets at 25°C.

**Device Performance and Environmental Response.** Raman spectroscopy (Figure S1)

confirms the monolayer quality and structural integrity of the graphene, displaying clear G (1589 cm$^{-1}$) and 2D (2680 cm$^{-1}$) peaks. The device's output performance critically depends on the interplate spacing. When silicon and graphene are in direct contact, electrical output is negligible. Systematic measurements at spacings of 100, 200, 300, and 400 μm reveal maximum output voltage (~0.35 V) and current (Figure S2) at 100 μm spacing, with progressive decrease at larger gaps (Figure 3a).

Device stability was validated through continuous 12-hour operation, demonstrating consistent performance (Figure 3b). Temperature-dependent measurements reveal a nonmonotonic voltage response: initial increase followed by decrease with rising temperature (Figure 3c). This trend arises from synergistic effects between water molecular polarity and evaporation kinetics. As shown in Table S1, while molecular polarity gradually declines at elevated temperatures, evaporation rate exhibits significant enhancement. To evaluate industrial potential, solar-integrated device testing under varying illumination revealed enhanced outputs of 0.75 V voltage and 1.05 μA current at 50 W/m² solar intensity (Figure 3d), attributed to synergistic photothermal acceleration of evaporation kinetics and photo-enhanced polarization of interfacial water molecules, collectively elevating device performance.

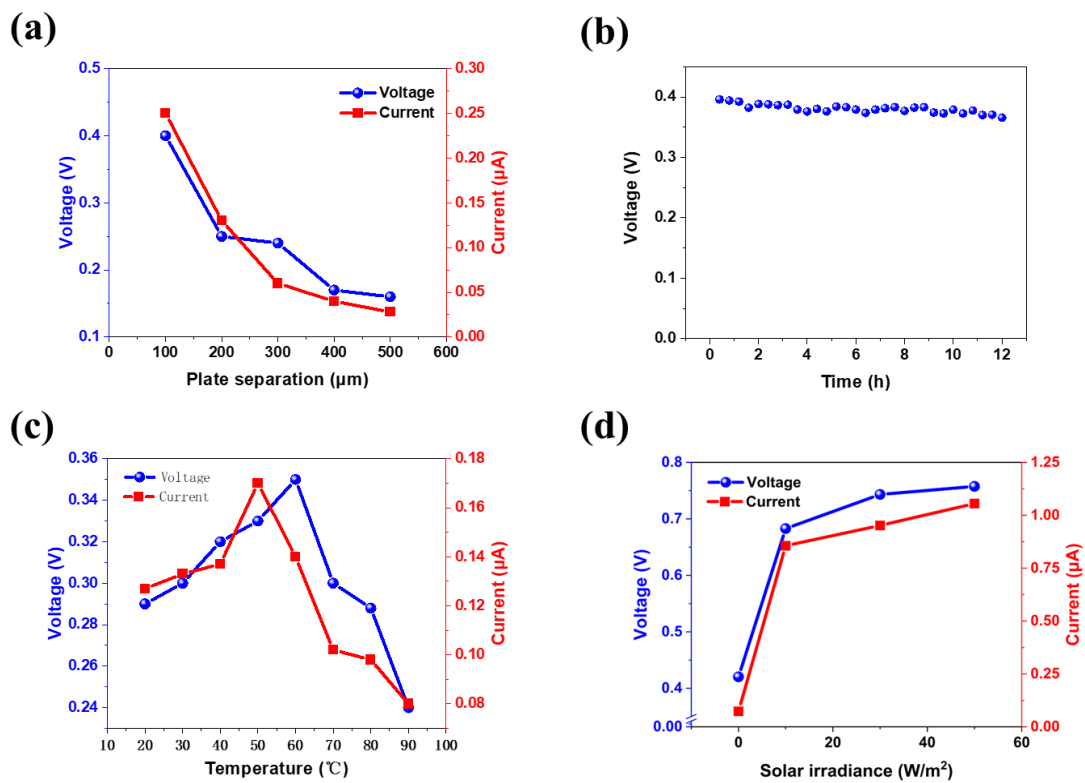

**Fig. 3 | Performance characterization of the graphene-water-silicon dynamic diode device.** (a) Dependence of Voltage Output on Interfacial Separation in Silicon-

Supported Graphene. (b) Long-term voltage stability tested over 12h continuous operation. (c) Open-circuit voltage dependence on ambient temperature (20-90°C). (d) Current and voltage output characteristics under varying solar irradiance levels.

**Demonstration and Application of a Modular Floating Generator.** Building upon the established experimental and theoretical framework, we engineered a floating generator prototype featuring a universal graphene–liquid–semiconductor architecture (Figure 4a). The 2 cm × 1 cm graphene–water–silicon device, integrated with an ethylene-vinyl acetate (EVA) foam flotation platform, comprises three functional strata: a bottom n-type silicon layer with 100-nm sputtered Ti/Au electrodes, a 100 μm-thick incorporated quartz spacers defining a 2 cm × 1 cm × 100 μm active cavity, and a top graphene membrane. This configuration enables stable operation on water surfaces through wave-induced cavity replenishment.

To expand application scenarios, we designed a vertically immersed device structure where the bottom section contacts water. Through capillary action, water wicks upward along the interlayer channel, generating continuously increasing voltage that stabilizes upon saturation. Four such vertically configured units connected in series (Figure 4c) collectively deliver ~1.2 V output (Figure 4d).

The combination of sustained voltage generation, scalable architecture, and compatibility with aqueous environments positions this technology as a promising candidate for powering floating IoT nodes, marine sensing platforms, and other self-sustained water-based electronic systems.

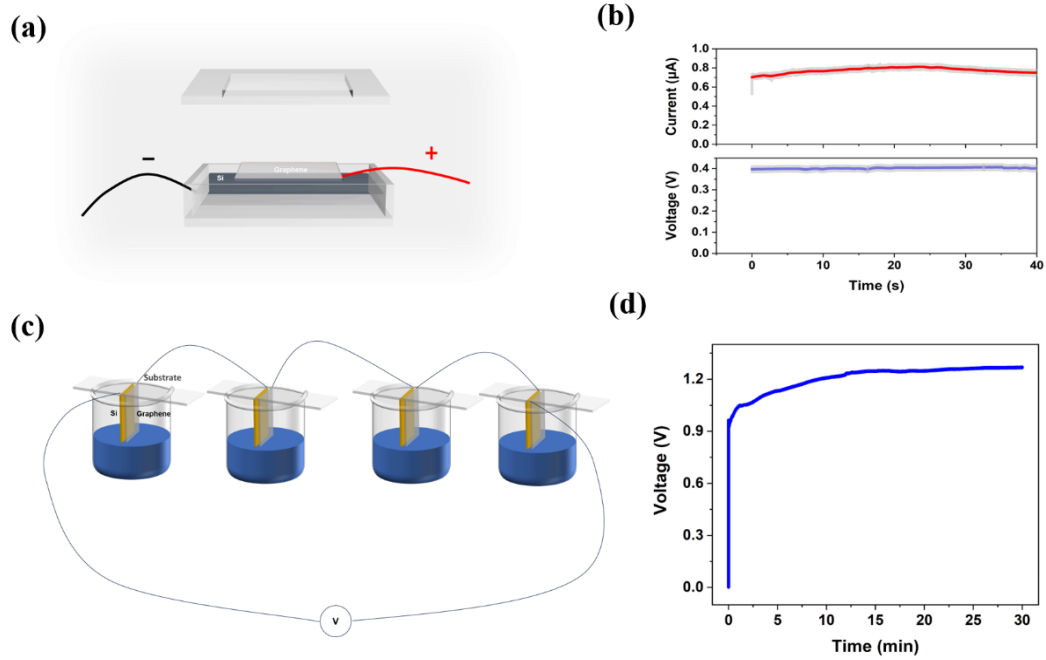

**Fig. 4 | Application demonstration of the graphene-water-silicon dynamic diode driven by water evaporation.** (a) Schematic illustration of the floating device architecture. (b) Continuous current (upper panel) and voltage (lower panel) outputs, with red (blue) curves representing smoothed current (voltage) profiles. (c) Schematic representation of four serially connected devices. (d) Output voltage characteristics of the quadruple-series configuration.

## Conclusion

In summary, we have demonstrated a novel DC generation mechanism through dynamically polarized graphene-water-semiconductor interfaces enabled by evaporation. The optimized device achieves sustained outputs of 0.35 V, with voltage polarity and magnitude governed by the Fermi level difference ($\Delta E_F$) between constituent materials. Systematic studies confirm the universality of this principle across metal-polar liquid-semiconductor architectures, exemplified by a packaged four-cell series generator delivering 1.2 V continuous output. Key advances over conventional water-based energy harvesters include : Directional Current Control: Voltage polarity programmable through $\Delta E_F$ engineering. The compatibility with diverse semiconductors and polar liquids paves an exciting path for potential future applications. Environmental Robustness: Encapsulated design ensures stable operation under ambient conditions. This work establishes interfacial polarization dynamics as a new paradigm for sustainable energy harvesting, offering a viable pathway toward self-powered aqueous electronics. The fundamental insights into non-electrolyte molecular orientation effects and the developed structure-property relationships provide a framework for optimizing liquid-solid energy conversion systems.